\documentclass[5p]{elsarticle}
\usepackage{hyperref}
\usepackage{booktabs}
\usepackage{amssymb}     % packages called in main file, e.g. gstr-template.tex
\usepackage{marvosym}    % symbols for EURO
\usepackage{upgreek}     % for upright greek letters in units
\usepackage{amsmath}
\usepackage[T1]{fontenc}

\newcommand{\gerda}       {\textsc{Gerda}} 
\newcommand{\majorana}    {\textsc{Majorana}}

\newcommand{\mubq}        {{$\upmu$Bq}}

\newcommand{\mus}         {{$\upmu$s}}

\newcommand{\ctsroiton}   {cts/(RoI$\cdot$ton$\cdot$yr)}
\newcommand{\qbb}         {{$Q_{\beta\beta}$}}

\newcommand{\etal}        {{\textit{et al.}}}
\newcommand{\kgyr}        {{kg$\cdot$yr}}

\journal{Astroparticle Physics}
\begin{document} 

\begin{frontmatter}
\title{Limits on uranium and thorium bulk content in \mbox{\sc{Gerda}} Phase~I
  detectors}

%  authors per affiliation:
\author{\mbox{\protect{\sc Gerda}} collaboration\fnref{aalab}
\corref{mycorrespondingauthor}%\fnref{aalab}
\cortext[mycorrespondingauthor]{Correspondence: {gerda-eb@mpi-hd.mpg.de}}}
\fntext[aalab]{Laboratori Nazionali del Gran Sasso, Assergi, Italy}
%\ead{gerda-eb@mpi-hd.mpg.de}
%
\author[addlngs]{~\\[3mm]M.~Agostini}
\author[adddre]{M.~Allardt}
\author[addku]{A.M.~Bakalyarov}
\author[addlngs]{M.~Balata}
\author[addinr]{I.~Barabanov}
\author[adduzh]{L.~Baudis}
\author[addhd]{C.~Bauer}
\author[addmpp]{N.~Becerici-Schmidt}
\author[addmibf,addmibinfn]{E.~Bellotti}
\author[additep,addinr]{S.~Belogurov}
% ...
\address[addlngs]{INFN Laboratori Nazionali del Gran Sasso, LNGS, and Gran
  Sasso Science Institute, GSSI, Assergi, Italy}
\address[addals]{INFN Laboratori Nazionali del Sud, Catania, Italy}
\address[addcra]{Institute of Physics, Jagiellonian University, Cracow, Poland}
\address[adddre]{Institut f{\"u}r Kern- und Teilchenphysik, Technische
  Universit{\"a}t Dresden, Dresden, Germany}
\address[addjinr]{Joint Institute for Nuclear Research, Dubna, Russia}
\address[addgeel]{European Commission, JRC-Geel, Geel, Belgium}
\address[addhd]{Max-Planck-Institut f{\"u}r Kernphysik, Heidelberg, Germany}
\address[addmibf]{Dipartimento di Fisica, Universit{\`a} Milano Bicocca,
     Milan, Italy}
\address[addmibinfn]{INFN Milano Bicocca, Milan, Italy}
\address[adduminfn]{Dipartimento di Fisica, Universit{\`a} degli Studi di
  Milano e INFN Milano,  Milan, Italy}
\address[addinr]{Institute for Nuclear Research of the Russian Academy of
  Sciences, Moscow, Russia}
\address[additep]{Institute for Theoretical and Experimental Physics,
    Moscow, Russia}
\address[addku]{National Research Centre ``Kurchatov Institute'', Moscow,
  Russia}
\address[addmpp]{Max-Planck-Institut f{\"ur} Physik, Munich, Germany}
\address[addtum]{Physik Department and Excellence Cluster Universe,
    Technische  Universit{\"a}t M{\"u}nchen, Munich, Germany}
\address[addpduni]{Dipartimento di Fisica e Astronomia dell{`}Universit{\`a}
  di Padova, Padova, Italy}
\address[addpdinfn]{INFN  Padova, Padova, Italy}
\address[addtue]{Physikalisches Institut, Eberhard-Karls Universit{\"a}t
  T{\"u}bingen, Germany}
\address[adduzh]{Physik Institut der Universit{\"a}t Z{\"u}rich, Z{\"u}rich,
    Switzerland}
\begin{abstract}
Internal contaminations of $^{238}$U, $^{235}$U and $^{232}$Th in the bulk of
high purity germanium detectors are potential backgrounds for experiments
searching for neutrinoless double beta decay of $^{76}$Ge.  The data from
\gerda\ Phase~I have been analyzed for alpha events from the decay chain of
these contaminations by looking for full decay chains and for time
correlations between successive decays in the same detector. No candidate
events for a full chain have been found.  Upper limits on the activities in
the range of a few nBq/kg for $^{226}$Ra, $^{227}$Ac and $^{228}$Th, the
long-lived daughter nuclides of $^{238}$U, $^{235}$U and $^{232}$Th,
respectively, have been derived. With these upper limits a background index in
the energy region of interest from $^{226}$Ra and $^{228}$Th contamination is
estimated which satisfies the prerequisites of a future ton scale germanium
double beta decay experiment.
\end{abstract}
%end of abstract
\end{frontmatter}
%
%%%%%%%%%%%%%%%%%%%%%%%%%%%%%%%%%%%%%%%%%%%%%%%%%%%%%%%%%%%%%%%%%%%%%%%%%%%%%%%
\section{Introduction} 
\label{sec:introduction} 
The unknown nature and the mass of the neutrino are among the biggest unsolved
mysteries of modern particle physics. Observation of neutrinoless double beta
(0$\nu\beta\beta$) decay could shed light onto these questions. The
experimental search for neutrinoless 0$\nu\beta\beta$ decay requires ultra
sensitive detectors with a very low background rate like High Purity Germanium
(HPGe) detectors.  The \gerda\ experiment~\cite{GERDA} has set the strongest
lower bound for the half-life of $0\nu\beta\beta$ decay of
$^{76}$Ge~\cite{PhysRevLett.111.122503,gerdaII}.  Further generations
of experiments with larger target masses have started or are in the planning
stage~\cite{majodemo}.

In order to improve the sensitivity of HPGe detector based experiments, an
increase of detector mass in combination with a further reduction of the
radioactive background is necessary. In the next generation of experiments the
detector mass could amount to about a ton of detector material while a
background rate below $\approx$~0.1\,\ctsroiton\ in the energy region of
interest (RoI) will be necessary. This value is more than one order of
magnitude lower than the level presently reached in
\gerda\ Phase~II~\cite{gerdaII}. In order to achieve such a low background
level, it is necessary to investigate and eventually reduce all possible
background components.

\begin{figure}[h]
\begin{center}
\includegraphics[width=0.99\columnwidth]{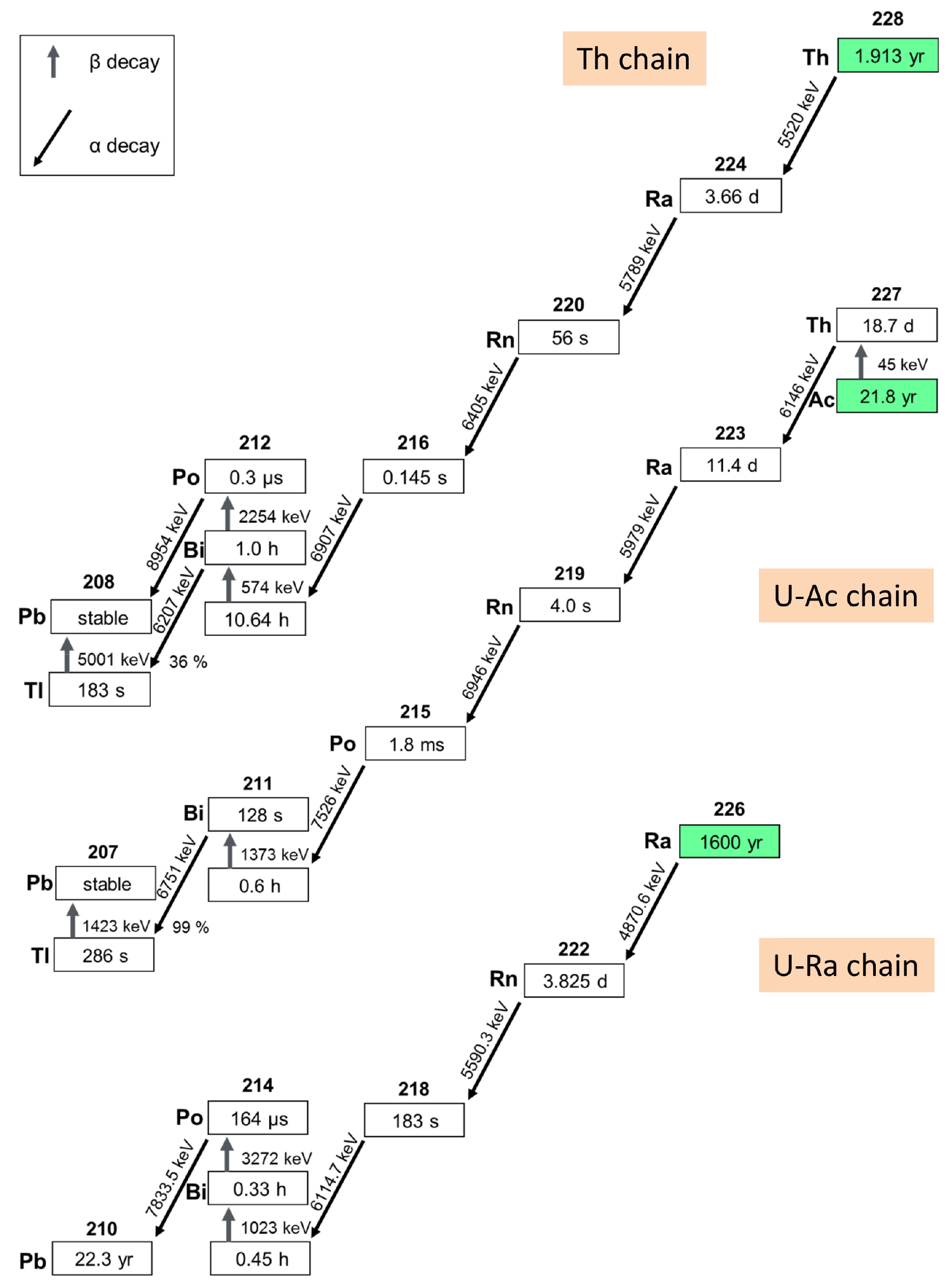}
\caption{  \label{fig:chains}
Overview of the relevant isotopes in the three natural decay chains
$^{232}$Th (Th chain), $^{235}$U (U-Ac chain) and $^{238}$U (U-Ra
chain). Energies, half-lives and branching ratios are given
}
\end{center}
\end{figure}

An important background contribution to all rare event searches is the
natural radioactivity from isotopes of the $^{238}$U, $^{235}$U and $^{232}$Th
decay chains. Fig.~\ref{fig:chains} depicts the relevant and accessible part
of the chains.  
Background in the RoI can among others be induced by several $\gamma$ lines
originating from the decays of $^{214}$Bi and $^{208}$Tl that are above the
$Q$-value of $0\nu\beta\beta$ decay in $^{76}$Ge, with \qbb~=~2039\,keV. The
most prominent one in general is the 2614.5\,keV line in $^{208}$Pb following
the $^{208}$Tl $\beta$ decay. These long-lived uranium and thorium isotopes
are omnipresent in the earth's crust~\cite{HansWedepohl19951217} and are to be
expected in any natural material and surrounding. The purification process of
germanium during production of HPGe crystals efficiently removes uranium and
thorium. However, further processing steps in detector production and setup
follow until the final deployment within the experiment. This might bring back
some fraction of radioactivity, mostly on the surface. A tiny contamination of
the bulk HPGe with these elements can deteriorate the sensitivity of HPGe
experiments considerably. Simulations suggest that contaminations of
$\approx$$0.1$\,\mubq/kg for $^{238}$U and $^{232}$Th account for an expected
background rate of 1\,\ctsroiton\ and 0.2\,\ctsroiton, respectively~\cite[and
  refs. therein]{KevinPhD} when a RoI of 4~keV around \qbb\ is assumed. Thus,
for future ton-scale experiments the bulk contamination of HPGe can be a
potential background source. This implies that the purity of HPGe has to be
controlled to a level lower than 10\,nBq/kg.

In this paper limits for the bulk contamination of HPGe detectors by isotopes
of the decay chains $^{238}$U, $^{235}$U and $^{232}$Th are given that are
obtained from \gerda\ Phase~I data.  The spectrum of every detector was
searched for a sequence of $\alpha$ decays occuring in the decay chains. In
most cases at least one of the decays was missing. If candidates for all
$\alpha$ decays exist the compartibility of the time distribution of the
events with the expectation from the half lives is checked.  The limits
derived from \gerda\ Phase~I data show for the first time that the bulk
contamination of HPGe can be controlled to a level of a few nBq/kg for the
isotopes $^{228}$Th, $^{227}$Ac and $^{226}$Ra resulting from primordial
$^{238}$U, $^{235}$U and $^{232}$Th.  This demonstrates that bulk
contamination of HPGe with these isotopes is unlikely to limit a future
ton-scale HPGe experiment for the search of $0\nu\beta\beta$ decay.

%%%%%%%%%%%%%%%%%%%%%%%%%%%%%%%%%%%%%%%%%%%%%%%%%%%%%%%%%%%%%%%%%%%%%%%%%%%%%%%%
\section{Experiment and data sets} 
\label{sec:experiment_data_sets} 

The \gerda\ experiment consists of an array of germanium detectors submerged
into a cryostat filled with liquid argon (LAr). The cryostat has a diameter of
4\,m and is placed inside a water tank with a diameter of 10\,m. In Phase~I of
the experiment, eight refurbished semi-coaxial HPGe detectors from the
Heidelberg-Moscow~\cite{HdM} and \textsc{Igex}~\cite{PhysRevD.65.092007}
experiments made from germanium isotopically enriched in $^{76}$Ge were
deployed together with other semi-coaxial detectors with natural
abundance~\cite{gtf}. Some detectors had been replaced by five enriched Broad
Energy Germanium (BEGe) detectors~\cite{seven}.  Details of the
\gerda\ experimental setup can be found in Ref.~\cite{GERDA}.

For the current analysis, all data recorded in the period between December 2,
2011 and September 5, 2013 inclusive those of the natural detectors were used
which expands slightly the data set of Ref.~\cite{PhysRevLett.111.122503}
to 26.14\,kg$\cdot$\,yr.  For the calibration of the energy scale \gerda\ uses
$^{228}$Th sources.  Since the expected energy deposition of an internal
$\alpha$ background is above the highest energetic calibration peak at
2.6\,MeV, data from calibration runs were also used in the analysis. While the
increase of exposure is not significant ($<$2\%), this extends the continuity
of data taking which in turn improves the detection efficiency of the decay
sequences (see sec.~\ref{sec:systematics}).  The longest live time for a
subset of detectors corresponds to 543.8 days (see
Table~\ref{tab:detectors}). Unphysical events due to electromagnetic noise or
discharges are identified and rejected by a set of quality cuts. These cuts
take into account parameters of the charge pulse, e.g. rise time and baseline
fluctuations, and are described in Ref.~\cite{1742-6596-368-1-012047}. Pile-up
events, in which two distinct pulses are recorded during the digitization time
window of 80\,\mus, are also discarded~\cite{BG}. Events in coincidence
with a muon trigger are rejected. The individual live times and masses of the
detectors used in this analysis are listed in Table~\ref{tab:detectors}.
\begin{table}[t]
\begin{center}
	\caption{\label{tab:detectors} 
          Detector masses and live times. The active mass of GTF112 and GTF45
          have been
          estimated from the average of those of the four ANG detectors
}
\begin{tabular}{cccc}
\shortstack{{detector}\\{~}\\{~}\\{~}} & \shortstack{{live time}\\{(days)}} &
 \shortstack{{total mass}\\{(kg)}} & \shortstack{{active mass}\\{(kg)}}\\
       \midrule ANG2  & 543.8 & 2.833 & 2.468$\pm$0.145~\ \ \\
		ANG3  & 543.8 & 2.391 & 2.070$\pm$0.136~\ \ \\
		ANG4  & 543.8 & 2.372 & 2.136$\pm$0.135~\ \ \\
		ANG5  & 543.8 & 2.746 & 2.281$\pm$0.132~\ \ \\
		RG1   & 543.8 & 2.110 & 1.908$\pm$0.125~\ \ \\
		RG2   & 411.8 & 2.166 & 1.800$\pm$0.115~\ \ \\
		GD32B & 282.4 & 0.717 & 0.638$\pm$0.113~\ \ \\
		GD32C & 305.9 & 0.743 & 0.677$\pm$0.019~\ \ \\
		GD32D & 286.3 & 0.723 & 0.667$\pm$0.022~\ \ \\
		GD35B & 305.9 & 0.812 & 0.742$\pm$0.019~\ \ \\
		GTF32  & 154.7 & 2.321 & 2.251$\pm$0.116~\ \ \\
                GTF45 & 154.7 & 2.312 &2.242$\pm$0.150~\ \ \\
		GTF112& 543.8 & 2.965 & 2.571$\pm$0.150~\ \ \\
	\end{tabular} 
\end{center}
%        $^\star$ For the detectors GTF112 and GTF45 no active volume was
%        determined by scans with $\gamma$s and thus it was estimated from
%        measurements with the other semi-coaxial detectors.
\end{table}

Neither the emitted $\alpha$ particle, nor the recoiling nucleus, carrying
typically $\approx120$\,keV, have mean free paths longer than a few
$\upmu$m. Due to this reason only events with an energy deposited in a single
detector, i.e. with multiplicity one, are selected without significant loss of
exposure.

When a potential parent decay resulting from $^{228}$Th, $^{227}$Ac or
$^{226}$Ra or the subsequent daughter nuclei is found, the following time
window of ten daughter half-lives is examined for the presence of a possible
daughter decay. This search is performed for all $\alpha$ decays occuring in
the sub-decay chains shown in Table~\ref{tab:Ewindows}. The probability for
the daughter nucleus to decay within this time interval is higher than
99.9\,\%. The energy intervals where this search is performed are:
\begin{align}
\label{eq:windows}
E_{\max}&=Q_\alpha+100\,\text{keV,}\nonumber\\
E_{\min}&=E_{\alpha,\min}-100\,\text{keV,} 
\end{align}
where $Q_\alpha$ is the $Q$-value of the decay and $E_{\alpha,\min}$ is the
(lowest) energy of the $\alpha$ particle that populates the highest excited
level of the daughter nuclei with branching ratios above 2.5\,\%. Branching
ratios below 2.5\,\% have been disregarded leading to a reduction of
efficiency of less than 4\,\% for the $^{227}$Ac sub-decay chain and less than
0.8\,\% for the $^{226}$Ra and $^{228}$Th chains.  The energy window is
enlarged by $\pm100$\,keV to account abundantly for the systematic uncertainty
in energy calibration from possible preamplifier gain non-linearity for
energies above 4.5\,MeV~\cite{BG}.  While the uncertainty is estimated to be
less than 50~keV at 6~MeV, it was decided to use the larger window size as it
does not change the result thanks to the quasi background free analysis while
being maximally conservative. By using the minimum $\alpha$ energy for
determination of the lower bound on the energy window, also the uncertainty on
the quenching factor of the recoiling nucleus is taken into account,
effectively allowing for a 100\,\% ionization quenching of the recoil nucleus.
The following analysis makes use only of part of the decay chains looking for
$\alpha$ decays with branching ratios of $>$96\,\% as shown in
Fig.~\ref{fig:chains} (see Table~\ref{tab:Ewindows}).

As the detection efficiency of $\alpha$ particles inside germanium detectors
is basically 100\,\%, from the non observa-

\begin{table}[h]
\begin{center}
\caption{\label{tab:Ewindows}
          Sub-decay chains considered in the search and corresponding energy
          windows where the search for $\alpha$ decay candidates is
          performed. The intervals are calculated according to
          Eq.~\eqref{eq:windows} with values from~\cite{isotopes}. The last
          column gives the averaged probability $\varepsilon$ to observe the
          decay of the daughter nucleus  (or the next $\alpha$ decay in the
          sequence in case the daughter nucleus is not an $\alpha$ emitting
          isotope) following the specific $\alpha$ decays listed for the
          detectors that have candidate events for all decays in the sub-decay
          chain
}
\begin{tabular}{cccr}
   chain&isotope & \shortstack{energy window (keV)}& $\varepsilon$\\
\midrule
        U-Ra & $^{226}$Ra & 4501 - 4971 & 1.000\\
%             & $^{222}$Rn & 5389 - 5690 & 0.847\\
             & $^{222}$Rn & 5389 - 5690 & 0.919\\
             & $^{218}$Po & 5902 - 6215 & 0.999\\%[3mm]
%             & $^{218}$Po & 5902 - 6215 & 0.909\\%[3mm]
\midrule                                       
        U-Ac & $^{227}$Th & 5601 - 6246 & 1.000\\
             & $^{223}$Ra & 5440 - 6079 & 0.878\\
%             & $^{223}$Ra & 5440 - 6079 & 0.862\\
             & $^{219}$Rn & 6325 - 7046 & 1.000\\
             & $^{215}$Po & 7286 - 7626 & 1.000\\
             & $^{211}$Bi & 6178 - 6851 & 0.995\\
\midrule                                       
          Th & $^{228}$Th & 5240 - 5620 & 1.000\\
             & $^{224}$Ra & 5340 - 5889 & 0.921\\
%             & $^{224}$Ra & 5340 - 5889 & 0.876\\
             & $^{220}$Rn & 6188 - 6505 & 0.993\\
             & $^{216}$Po & 6678 - 7007 & 1.000
	\end{tabular}
\end{center}
\end{table}

\noindent \-tion of an intermediate decay within
a chain it can be
followed with high confidence that the mother nucleus of the
observed decay can not have been inside the bulk of the material.  The
$\alpha$ decays and the corresponding energy windows to which the search is
restricted are listed in Table~\ref{tab:Ewindows}.

\begin{figure}[thb]
\begin{center}
 \includegraphics[width= 0.99\columnwidth]{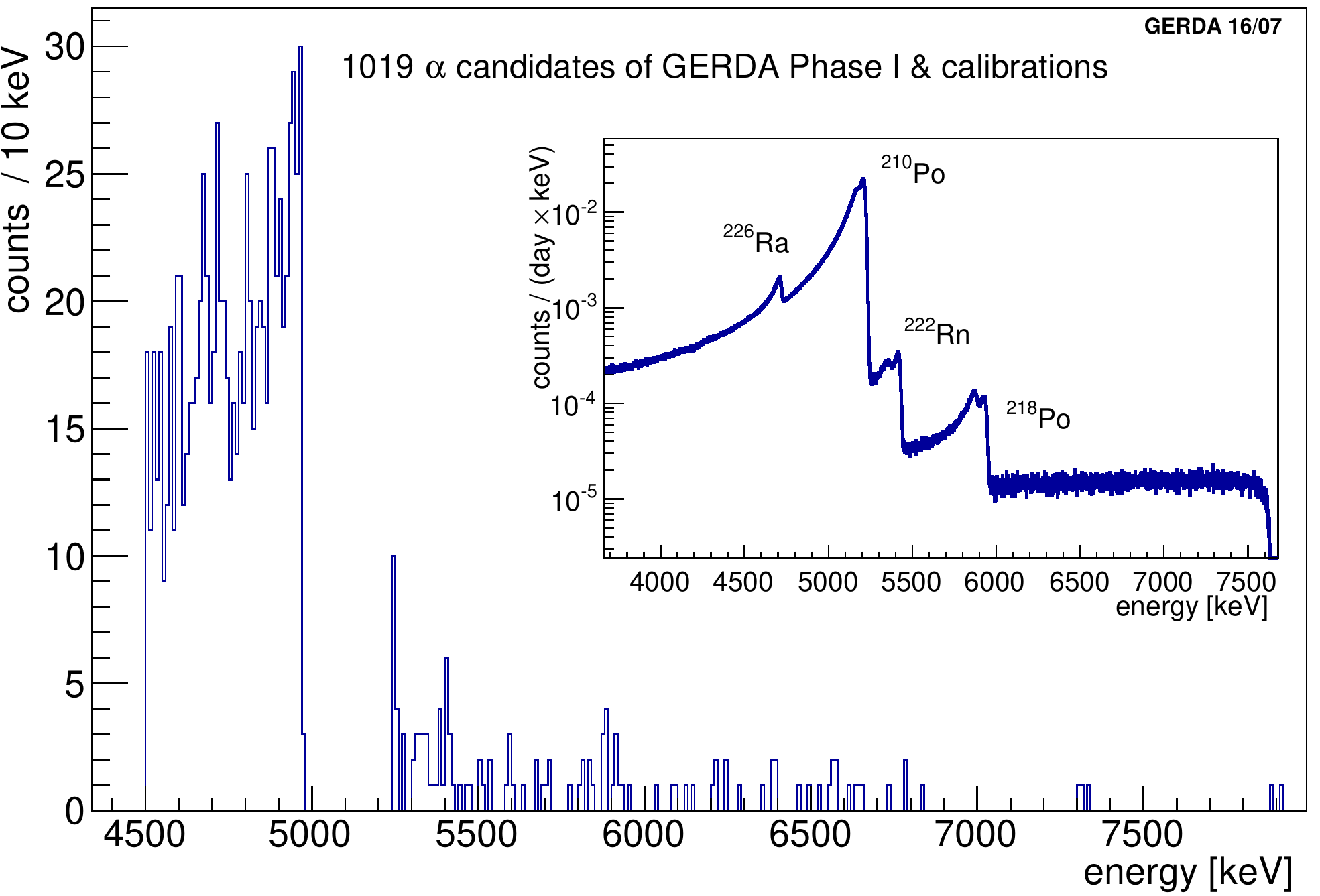}%
  \caption{\label{fig:data}
         Energy spectrum of $\alpha$ decay candidates from the uranium and
         thorium series for the whole data-taking period considered in this
         analysis. For comparison the inset shows the prediction by the
         background model for the sum of the enriched semi-coaxial detectors.
}
\end{center}
\end{figure}

The energy spectrum of all candidate $\alpha$ events without time coincidence
cut for all detectors and for the complete data-taking period after selection
based on Eq.~\eqref{eq:windows} is shown in Fig.~\ref{fig:data}. The lowest
energy of interest given in Table~\ref{tab:Ewindows} is 4501\,keV which
defines the range in this figure. The gap in the spectrum appears due to the
gap in the allowed energy ranges (see Table~\ref{tab:Ewindows}).

The contamination is calculated as the mass fraction $C$, i.e. the mass $M$ of
the isotope under consideration over the active mass of the detector
$m_\text{det}$:
\begin{equation}
	\label{eq:contamination}
C=\frac{M}{m_\text{det}}=\frac{\nu\ m_A}{N_A\ m_\text{det}
  (1-e^{-\lambda \Delta t}) \varepsilon}\simeq \frac{\nu\ m_A}{N_A\ \lambda\ 
         \varepsilon}\  {\cal E}\,\ , 
\end{equation}
where $N_A$ is Avogadro's number, $\nu$ is the number of detected events or
the corresponding upper limit in case of no observed signal, $m_A$ is the
molar mass of the contaminating isotope, $\lambda$ is the decay constant,
${\cal E}=m_\text{det}\cdot\Delta t$ is the exposure.  The last approximation
in Eq.~\ref{eq:contamination} can be used when the observation time is much
shorter than the parent nuclide half-life ( $^{228}$Th, $^{227}$Ac and
$^{226}$Ra), which always holds true for this analysis. The efficiency,
$\varepsilon$, accounts for detector dead times and the probability that
subsequent events in a decay chain happen outside the active volume. It is
estimated in sec.~\ref{sec:systematics} an given in Table~\ref{tab:Ewindows}.

\section{Background estimation for time coincidence signature}
\label{sec:BG}

The energy spectrum at the relevant energies for this analysis is dominated by
$\alpha$ decays of $^{210}$Po, $^{226}$Ra, $^{222}$Rn and $^{218}$Po on the p+
surfaces of the detectors~\cite{BG}. This surface contamination can be the
source of successive decays that mimic the time and energy signature of signal
$\alpha$ decays within the bulk. In the following we assume constant decay
rates in the energy windows (see Table~\ref{tab:Ewindows}) taken from the
$\alpha$ background model developed in Ref.~\cite{BG}.  In reality this is not
accurate for the $^{210}$Po part of the contamination.  The analysis is,
however, not sensitive to the details of the alpha background model.  The
probability density that a background event in the energy range expected for
the daughter isotope $(i+1)$ will take place stochastically in time $t$ after
a background event $i$ in the energy range corresponding to the mother
isotope, is given by the product of the random time interval distribution of
background events in the energy range of the daughter isotope $(i+1)$:
%chain~\cite{knoll2010radiation,ErrorsPoorStat}:
\begin{equation}
	\label{eq:prob}
\frac{dp}{dt}=r_{i+1}\,\,e^{-r_{i+1}t}\,\, , 
\end{equation}
where $r_i$ is the mean background rate in the energy window of decay $i$. The
integral of Eq.~\eqref{eq:prob} from zero to $\Delta t_{i,i+1}$ is the
probability to randomly observe the decay sequence $i$,$(i+1)$ within the time
window $\Delta t_{i,i+1}$. The number of expected random coincidences from
background events to the present chain analysis, i.e. random decay sequences
that mimic the signal decay chain during the life time $T$ is then given by:
\begin{equation}
	\label{eq:num}
n_b=r_1\,T\,
%\frac{\displaystyle\prod_{i=1}^{K}r_i}%
%       {{\left(\displaystyle\sum_{i=1}^{K}r_i\right)}^{K-1}}\,
 \displaystyle\prod_{i=1}^{K-1}%
   \Big[1-e^{-r_{i-1}\,\Delta t_{i,i+1}}\Big]\,\,\ , 
\end{equation}
where $K$ is the number of decays in the sequence and $\Delta t_{i,i+1}$ are     
maximum time limits between two subsequent decays $i$ and $(i+1)$, i.e. ten
half-lives for the present analysis.

The $\alpha$ background model~\cite{BG} was used to estimate the rates $r_i$
in Eqs.~\eqref{eq:prob} and \eqref{eq:num}.  For a live time of 543.8 days,
the number of stochastic coincidences, $n_b$, is $n_b<5\cdot10^{-15}$ for the
$^{227}$Th, $n_b<1\cdot10^{-13}$ for the $^{228}$Th and $n_b<0.94$ for the
$^{226}$Ra chain. The very low background expectations for the $^{227}$Ac and
$^{228}$Th sub-chains appear due to the exponential suppression of the
probability for stochastic coincidence with smallness of the time windows
between two subsequent decays. Both sub-chains have two subsequent decays with
time windows of the order of seconds or less.

An additional background component is due to $^{228}$Th, $^{227}$Ac or
$^{226}$Ra contamination on the p+ surfaces of the detectors. Such surface
contaminations can lead to the same time signature while not being related to
bulk contamination of the HPGe material itself.  Considering a mother nucleus
at rest on the p+ surface, the $\alpha$ particle and the daughter nucleus will
have momenta in opposite directions. Hence, if the $\alpha$ particle is
detected through energy deposition in the active volume, the daughter nucleus
will likely recoil away from the p+ surface. In order to observe the decay of
the daughter nucleus, its momentum needs to be pointing towards the active
volume. However, in this case, the $\alpha$ particle from the preceeding decay
of the mother nucleus will be traveling away from the active volume and likely
remain undetected. Therefore, based on simple kinematic arguments, it is
unlikely that all $\alpha$ decays from p+ surface contaminations will be
observed. This argument is supported by the observation that for the best fit
background model in Phase~I, the number of events from an initial $^{226}$Ra
contamination on the p+ surface is roughly halved for each subsequent
decay~\cite{BG}. In order to quantify this background for the $^{226}$Ra sub-chain 10$^5$ decay chains originating from $^{226}$Ra nuclei in the dead layer
were simulated using \textsc{MaGe} \cite{mage}.  From these simulations it is
derived that the percentage of $^{226}$Ra decays that will produce three
subsequent $\alpha$ decays depositing energy in the active volume is expected
to be 0.07\,\%. For 50.5 $^{226}$Ra decays on the p+ surface as derived from
the $\alpha$ background analysis~\cite{BG}, the expected number of triple
coincidences is less than 0.04.

\section{Efficiency estimation}\label{sec:systematics}

When nuclei of parent isotopes decay in the active volume of a detector, all
subsequent decays of daughter nuclides in the related chain should be
observable as long as the measuring time is considerably longer than the
longest half-life in the chain. The probability that the daughter nucleus with
typically $\approx$\,120\,keV recoil energy after an $\alpha$ decay leaves the
active volume of the detector due to its recoil is negligible, as the range of
heavy ions with this recoil energy is only a few tens of nm. Hence, nuclei can
be kicked out of the active volume only from the outer $\approx$\,50\,nm of
the detector, corresponding to an active volume fraction of $\approx$10$^{-6}$
for detectors with $\approx$\,4\,cm radius and $\approx$\,10\,cm height.

For decays in the bulk of the detector without the emission of gamma particles
the efficency of deposition of the full energy in the detector released in the
decay is practically 100 \%.  In this case the full energy released in the
decay can be detected. Both the recoiling nucleus and the emitted $\alpha$
particle deposit their full energy with high probability in the active
volume. For the recoiling nucleus the same argument holds as above. An
$\alpha$ particle with a few MeV of kinetic energy has a stopping range of a
few tens of $\upmu$m~\cite{Ziegler2008}. Hence, the probability that it leaves
the active volume will be larger, but still negligible with respect to the
uncertainty on the active volume. Conserva\-ti\-ve\-ly assuming that only
$\alpha$ particles from the outer $\approx$\,50\,$\upmu$m shell of the active
volume of the detector escape from the active volume, its fraction is reduced
by approximately one permil.

If there are candidates for the decay of a mother nucleus of a given sub-decay
chain, when searching for time correlations between two or more events,
detector off times play a crucial role. As the subsequent decays could happen
during detector off-time, the exposure has to be corrected.  In order to
conservatively estimate this change in effective exposure due to detector
off-times, convolutions of the live time profile with the exponential decay
probability density function (pdf) of the expected decay are used. In
particular, for each 120 second time bin of each individual detector the
exponential pdf of the subsequent $\alpha$ decays have been convoluted with
the detector life time profiles following the given time bin.  This provides
the pdf of observing the daughter decay at that moment.  The integral over
this convoluted pdf then gives the reduction in effective life-time for the
given bin.  The average probabilities of observing the subsequent $\alpha$
decays within the decay chain are obtained by averaging over all time bins
during the life time of the individual detectors.  They are listed in the last
column of Table~\ref{tab:Ewindows}.  For detectors that have no candidate
decay of the mother isotope or that have zero candidates of one of the
isotopes of a sub-decay chain listed in Table~\ref{tab:Ewindows} no change in
effective exposure is taken into account.

The time averaged probability to observe all $\alpha$ decays in the chain
after a candidate of the mother isotope has been detected, is 0.920 for
$^{228}$Th, 0.874 for $^{227}$Ac and 0.918 for $^{226}$Ra for the detectors in
which there were candidate events for all decays in the sub-decay chain.
These values are used to obtain the effective exposure $\varepsilon\times{\cal
  E}$ in Eq.~\ref{eq:contamination} for these detectors.  Taking this loss in
efficiency and all systematic uncertainties into account the effective
exposure is reduced to 23.9\,kg$\cdot$yr for the $^{228}$Th sub-decay chains,
24.1\,kg$\cdot$yr for the $^{227}$Ac and 23.3\,kg$\cdot$yr for the $^{226}$Ra.

\section{Results}
In total there are 1019 candidate $\alpha$ events.
Table~\ref{tab:nr_candidate} shows the total number of $\alpha$-candidate
events in each individual detector during the whole corresponding life time
for each individual decay considered.  Most detectors have zero events for at
least one step in the individual sub-decay chains.  For these detectors it is
evident without further analysis that no complete chain was observed within
the life time.

Additionally a coincidence search was performed. No full decay chain could be
identified.

\begin{table*}[hbt]
\begin{center}
 \caption{\label{tab:nr_candidate}
          Number of 
         candidate events in the individual detectors.  Note that candidate
         events can appear for more than one decay due to overlap of the
         energy windows.
% The first three
 %        columns indicate the number of total events in the allowed energy
 %        windows. The last three columns shows the number of events
 %        with a double coincidence, either with the mother nucleus or .
}
\hspace*{-7mm}\begin{tabular}{c|ccccccccccccc}
&ANG2&ANG3&ANG4&ANG5&RG1&RG2&GD32B&GD32C&GD32D&GD35B&GTF32&GTF45&GTF112\\
\midrule
$^{226}$Ra & 142 & 265 & 45 & 92 & 30 & 35 & 6 & 10 & 5 & 8 & 45& 35 & 144\\
$^{222}$Rn & 3 & 7 & 1 & 0 & 0& 0  & 0 & 1 & 0 & 0 & 1 & 3 & 12 \\
$^{218}$Po & 1 & 2 & 3 & 0 & 1 & 0 & 0 & 0 & 0 & 0 & 2 & 0 & 6  \\

\midrule
$^{227}$Ac & 5 & 7 & 5 & 1 & 1 & 0 &  0 & 0 & 0 & 1 & 3 & 2 &14 \\
$^{223}$Ra & 7 & 7 & 4 & 1 & 0 & 0  & 0 & 1 & 0 & 1 & 2 & 2& 17\\
$^{219}$Rn & 2 & 3 & 3 & 2 & 0 & 0 & 1 & 0 & 0 & 0  & 0 & 5 & 5 \\
$^{215}$Po & 0 & 1 & 0 & 0 & 1 & 0 & 0 & 0 & 0 & 0  & 1 & 0 & 0  \\
$^{211}$Bi  & 2 & 4 & 4 & 2 & 0 & 0  & 1  & 0 & 0 & 0 & 1 & 6 & 7\\

\midrule
$^{228}$Th& 10 & 14 & 2 & 7 & 1 & 0  & 0 & 2 & 0 & 0  & 2 & 4 & 20\\
$^{224}$Ra& 9 &   13 & 3 & 2 & 0 & 0  & 0 & 1 & 0 & 0 & 2 & 3 & 21 \\
$^{220}$Rn& 1 &    1 & 3 & 2 & 0 & 0   & 0 & 0 & 0 & 0 & 1 & 2 & 3  \\
$^{216}$Po& 0 &    1 & 0 & 0 & 0 & 0    & 0 & 0 & 0 & 0 & 0 & 2 & 1 \\
\hline

\end{tabular}
\end{center}
\end{table*}

\begin{table*}[hbt]% [htb]
 \centering 
 \caption{\label{tab:candidate}
          Triple coincidence candidate for the Th chain in
          GTF112 starting with $^{228}$Th
}
\begin{tabular}{ccccc}
 \shortstack{candidate\phantom{(V)}\\event\phantom{(V)}} & \shortstack{recorded energy\\(keV)}  &
 \shortstack{$\Delta E$ ($Q_\alpha-E$) \\ (keV)}&
 \shortstack{time\\\phantom{(V)}} & \shortstack{time to next event\\(days/half-lives)}\\
		\midrule
 $^{228}$Th & 5400.51 & 119.49 & Dec 12 20:54:41 2011 & 23.1 / 6.3\\
 $^{228}$Th & 5381.52 & 138.48 & Dec 20 17:10:47 2011 & 15.3 / 4.2 \\
		\addlinespace[1ex]
 $^{224}$Ra & 5388.16 & 400.84 & Jan  4 23:43:15 2012 & 0.0001/0.2\\
		\addlinespace[1ex]
 $^{220}$Rn & 6283.16 & 121.84 & Jan  4 23:43:26 2012 &           \\
	\end{tabular}
\end{table*}

In total the 1019 candidate $\alpha$ events produced 304 potential $^{226}$Ra
-- $^{222}$Rn, 65 $^{227}$Th -- $^{223}$Ra, 42 $^{228}$Th -- $^{224}$Ra, 1
$^{224}$Ra -- $^{220}$Rn and 1 $^{222}$Rn -- $^{218}$Po coincidences.  These
numbers can be compared with expectations from Eq.~\eqref{eq:contamination}
using the background model: This simple estimation gives an expectation of
$\sim$~300 $^{226}$Ra -- $^{222}$Rn, $\sim$50 $^{228}$Th--$^{224}$Ra and
$\sim$50 $^{227}$Th--$^{223}$Ra stochastic coincidences.  Additionally there
is one potential triple coincidence for a part of the $^{228}$Th sub-chain
from $^{228}$Th to $^{220}$Rn. Note that the same event can be part of more
than one candidate coincidence.

It is interesting to have a closer look at the triple coincidence in GTF112.
Some information is detailed in Table~\ref{tab:candidate}.  The isotope
candidate, the recorded energy, the energy difference to $Q_\alpha$ and
time of the event are listed in this table. The last column shows the time
interval to the next candidate decay in the chain. There are two possible
candidate events for the parent nucleus. It is worth noting that all candidate
events have energies that differ from the Q-values of the corresponding decays
by $\approx120$\,keV. This is consistent with the expectation that only the
energy from the $\alpha$ particle, and not the recoiling daughter nucleus, is
detected as expected for surface events. In contrast, one should note that
the energy deposit of the $^{224}$Ra candidate in Table~\ref{tab:candidate} is
300 keV less than the most probable energy of the emitted $\alpha$ particle
being 5685.37\,keV, with a probability of 94.92\,\%~\cite{isotopes}. The
decay of $^{216}$Po is missing.

A mono-parametric pulse shape analysis method to identify events on the p+
surface has been developed~\cite{AgostPhD}. The discriminating parameter is
the rise time between 5\,\% and 50\,\% of the maximum pulse amplitude,
$t_{5-50}$. The events of Table~\ref{tab:candidate} have a $t_{5-50}$ of
$\approx110$\,ns. These values are below the cut threshold of 990\,ns as given
in Ref.~\cite{AgostPhD} indicating that they are consistent with the
expectation for surface events.

Having found no candidate for a full decay chain and conservatively assuming
zero background, the 90\,\% credibility interval (C.I.) Bayesian upper limit
for the contamination is 2.3 events, considering all $\alpha$ decays in the
chains from $^{227}$Ac to $^{207}$Pb and $^{228}$Th to $^{212}$Pb,
respectively.  In case of $^{226}$Ra, the expected number of background events
is 1 and the 90\% C.I. limit is 2.3 events.  Since the half-life of $^{227}$Th
is much smaller than the one for $^{227}$Ac, secular equilibrium can be
assumed, hence the limit on $^{227}$Ac is given instead.  The derived limits
and activities, taking into consideration the effective exposures and
uncertainties in the active mass of the detectors are shown in
Table~\ref{tab:limits} in g/g and nBq/kg.

The 90~\% C.I. upper limit on the $^{226}$Ra contamination is
$8.6\cdot10^{-23}$\,g/g. The limits on the concentration of $^{227}$Ac and
$^{228}$Th are $1.1\cdot10^{-24}$\,g/g and $1.0\cdot10^{-25}$\,g/g,
respectively.  The upper 90\% C.I. limit on the activity of $^{226}$Ra is
3.1\,nBq/kg, the limit on the activity of $^{227}$Ac is 3.0\,nBq/kg and the
limit on the activity of $^{228}$Th is 3.1\,nBq/kg corresponding to a rate of
less then 100 decays per year per ton of germanium.

Assuming that in a purification process the ratio of the isotopes
$^{228}$Th and $^{232}$Th stays constant and that since the deviation from
secular equilibrium more than 20 years

\begin{table}[hb!]
\begin{center}
\caption{\label{tab:limits}
            90 \% C.I. upper limits on the concentration of $^{226}$Ra,
            $^{227}$Ac and $^{228}$Th bulk contamination in \gerda\ Phase~I
            detectors, expressed in g/g are given in the middle
            column. The right column gives upper limits on the
            activities of the bulk contamination in \gerda\ Phase~I detectors,
            expressed in nBq/kg
}
\begin{tabular}{ccc}
  & contamination & activity\\
  & [g/g] & [nBq/kg]\\
\midrule
$^{226}$Ra &  8.6$\cdot$10$^{-23}$&3.1\\
$^{227}$Ac & 1.1$\cdot$10$^{-24}$&3.0\\
$^{228}$Th & 1.0$\cdot$10$^{-25}$&3.1\\
\midrule
\end{tabular}
\end{center}
\end{table}

\noindent have passed (which is a good assumption
for the HdM and IGEX semi-coaxial detectors), the 90\% C.I. upper limit on the
$^{232}$Th contamination~\cite{radon} is $\sim$4\,nBq/kg for $^{232}$Th.
Assuming secular equilibrium for the $^{238}$U and $^{232}$Th decay chains and
a RoI window of 4\,keV, the expected background index contributions are
$\approx0.05$ and $\approx0.01$\,\ctsroiton, respectively.  This
analysis shows that HPGe belongs among the most radiopure materials.  With the
same analysis with GERDA Phase II data with an exposure of 100~\kgyr\ we expect
an improvement of the sensitivity to bulk alphas by a factor of roughly~5,
i.e. a sub nBq/kg sensitivity.

\section{Conclusions}
A search for a possible presence of contamination, originating from the
uranium and thorium natural radioactivity series, in the bulk of HPGe
detectors in \gerda\ Phase~I has been performed. No complete decay chain was
observed. From the non-observation of full decay chains 90\,\% C.L. upper
limits were derived for the concentrations and activities of $^{228}$Th,
$^{227}$Ac and  $^{226}$Ra in \gerda\ Phase~I detectors.  A contamination of
detectors according to these limits would lead to a background index
0.1\,\ctsroiton in accordance with the requirements for a future ton scale
double beta decay experiment based on germanium detectors.

\section*{Acknowledgments}
The \gerda\ experiment is supported financially by the German Federal
Ministry for Education and Research (BMBF),
the German Research Foundation (DFG)
via the Excellence Cluster Universe, the Italian Istituto Nazionale
di Fisica Nucleare (INFN), the Max Planck Society (MPG),
the Polish National Science Center (NCN),
the Russian Foundation for Basic Research (RFBR),
and the Swiss National Science Foundation (SNF).
The institutions acknowledge also internal financial support.

%\bibliography{References}

\end{document}